\documentclass[sigconf]{acmart}

\usepackage{booktabs}
\usepackage{threeparttable,booktabs}
\usepackage{etoolbox}
\appto\TPTnoteSettings{\footnotesize}
\usepackage{siunitx}
\usepackage{tikz}
\usepackage{tablefootnote}
\usepackage{lipsum}
\usepackage{threeparttable}
\usepackage{amsfonts}
\usepackage{nicematrix,enumitem,booktabs}
\newcommand*{\priority}[1]{\begin{tikzpicture}[scale=0.15]%
    \draw (0,0) circle (1);
    \fill[fill opacity=0.5,fill=black] (0,0) -- (90:1) arc (90:90-#1*3.6:1) -- cycle;
    \end{tikzpicture}}
\AtBeginDocument{%
  \providecommand\BibTeX{{%
    \normalfont B\kern-0.5em{\scshape i\kern-0.25em b}\kern-0.8em\TeX}}}

\acmConference[Preprint Under review]{}{}{}
\begin{document}

\title{SoK: Privacy Preserving Machine Learning using Functional Encryption: Opportunities and Challenges}

 \author{Prajwal Panzade}
 \affiliation{
 \institution{Georgia State University}
   \city{Atlanta}
   \state{GA}
   \country{USA}
 }
 \email{ppanzade1@student.gsu.edu}
  \author{Daniel Takabi}
 \affiliation{
 \institution{Georgia State University}
   \city{Atlanta}
   \state{GA}
   \country{USA}
 }
 \email{takabi@gsu.edu}

\begin{abstract}
 With the advent of functional encryption, new possibilities for computation on encrypted data have arisen. Functional Encryption enables data owners to grant third-party access to perform specified computations without disclosing their inputs. It also provides computation results in plain, unlike Fully Homomorphic Encryption. \\ 
The ubiquitousness of machine learning has led to the collection of massive private data in the cloud computing environment. 
This raises potential privacy issues and the need for more private and secure computing solutions. Numerous efforts have been made in privacy-preserving machine learning (PPML) to address security and privacy concerns. There are approaches based on fully homomorphic encryption (FHE), secure multiparty computation (SMC), and, more recently, functional encryption (FE). 
However, FE-based PPML is still in its infancy and has not yet gotten much attention compared to FHE-based PPML approaches. \\ In this paper, we provide a systematization of PPML works based on FE summarizing state-of-the-art in the literature. We focus on Inner-product-FE and Quadratic-FE-based machine learning models for the PPML applications. We analyze the performance and usability of the available FE libraries and their applications to PPML. We also discuss potential directions for FE-based PPML approaches. To the best of our knowledge, this is the first work to systematize FE-based PPML approaches.
\end{abstract}




\maketitle

\section{Introduction}
\label{section:1}
Machine learning's expansion into domains such as computer vision, natural language processing, and speech processing has resulted in a plethora of incredible applications that have become an inseparable part of people's lives. More real-world machine learning applications today rely on a cloud computing environment following the concept of machine learning as a service (MLaaS) \cite{ribeiro2015mlaas}. More and more highly regulated enterprises and organizations, for example, banks, governments, insurances, and health, are migrating their data and machine learning services to the cloud. As a result of this advancement, there arises an increase in demand for secure and confidential computing solutions that preserve data and model privacy in ML applications relying on cloud-based machine learning. In light of this, researchers are paying close attention to privacy-preserving machine learning (PPML) \cite{hesamifard2018privacy, securenn, deepsecure}. The PPML's purpose is to deal with the problems related to data and model privacy during the training to deployment stages of machine learning.

\begin{figure*}[!ht]
\centering
\includegraphics[scale=0.5]{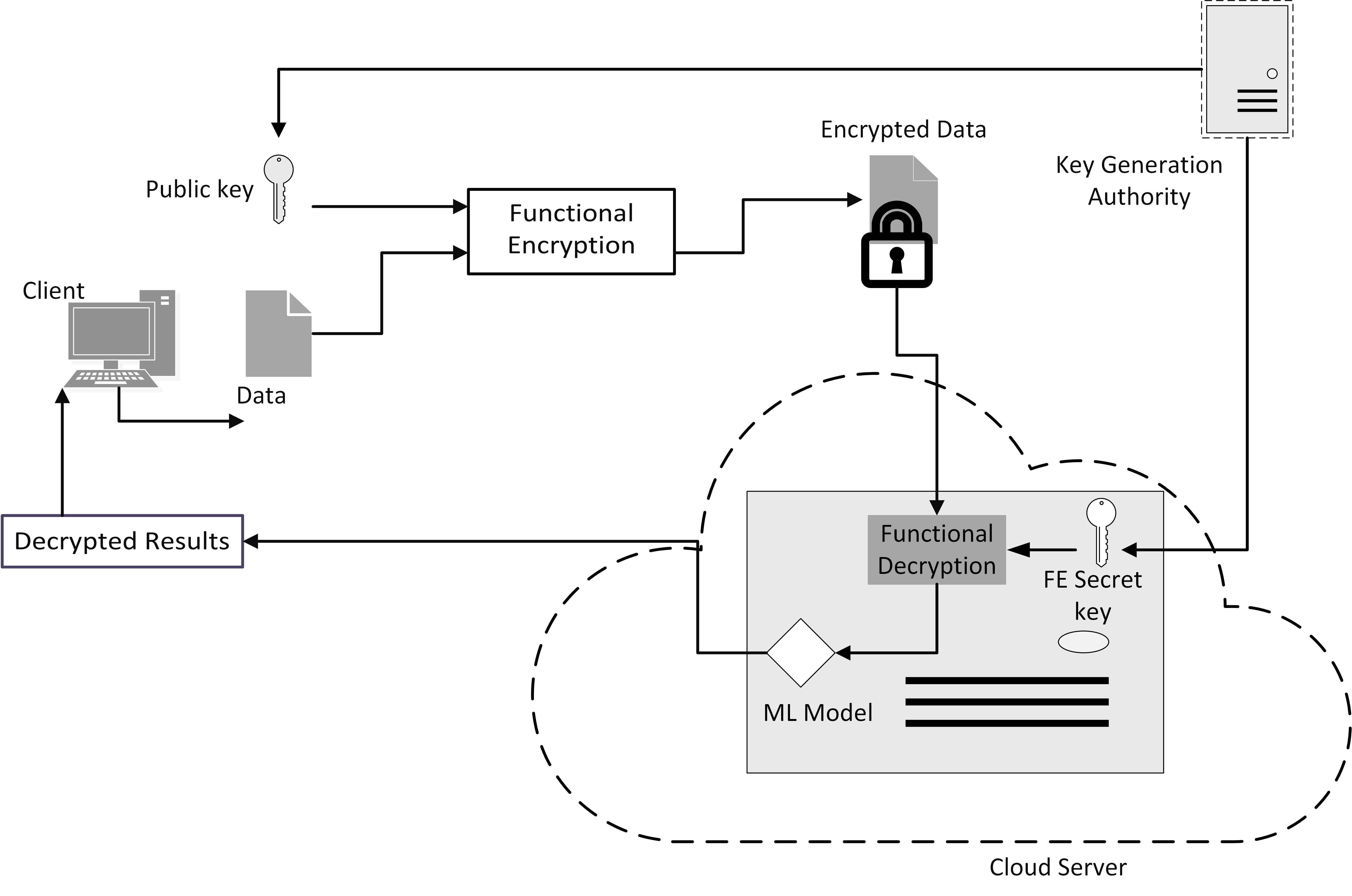}
\caption{\centering Overview of PPML using Functional Encryption
\label{fig:1}}
\end{figure*}
In the literature, fully homomorphic encryption (FHE) \cite{gentry2009fully}, secure multi-party computation (SMC) \cite{bogetoft2009secure}, federated learning \cite{mcmahan2017communication}, differential privacy \cite{dwork2006calibrating}, and trusted execution environments \cite{sabt2015trusted} are existing approaches to achieve PPML. In addition to this, functional encryption \cite{boneh2011functional} is also advancing day by day. Fully homomorphic encryption is an encryption standard that allows users to compute over encrypted data without decrypting the ciphertext itself. 
Secure multi-party computation allows many parties to jointly compute a function over their inputs while keeping their inputs private. On the other hand, federated learning is a machine learning strategy that allows multiple clients to train a model owned by a server while keeping the input data of each participant private. Differential privacy enables the quantification of an algorithm's level of privacy protection to the underlying sensitive dataset and facilitates machine learning model training on private data. A secure enclave, also known as a trusted execution environment (TEE), is a combination of software and hardware elements that creates an isolated environment in a system that ensures the security of applications running inside the TEE.  Functional encryption is a type of encryption comparable to fully homomorphic encryption in which computations can be performed on encrypted data. The key distinction between fully homomorphic and functional encryption is that fully homomorphic encryption produces ciphertext results, whereas functional encryption produces plaintext results.

There are primarily two ways to achieve PPML: Fully Homomorphic Encryption \cite{gentry2009fully} and Secure Multiparty Computation \cite{bogetoft2009secure}. Dowlin et al. proposed a method for transforming a trained neural network to a CryptoNet \cite{cryptonet}, that enables data owners to send homomorphically encrypted data to the server and receive an encrypted inference.
In another approach called CryptoDL, Hesamifard et al. \cite{cryptodl} proposed an FHE-based method for privacy-preserving inference on pre-trained convolutional neural networks. Al-Badawi et al. developed Private Fast Text (PrivFT)\cite{privft}, a method that utilizes FHE for privacy-preserving text classification. Li et al. introduced HomoPAI \cite{homopai}, an FHE-based machine learning platform built on top of PAI (Platform of Artificial Intelligence, an Alibaba Cloud product), enabling data owners from different organizations to store their encrypted data and perform collaborative machine learning. Jäschke et al. in \cite{unsup} proposed an approach to perform unsupervised machine learning where they implemented a K-means clustering algorithm using FHE. 

Graepel et al. in ML Confidential \cite{graepel2012ml} proposed a binary classification method using polynomial approximations and FHE. In SecureML \cite{secureml}, Zhang et al. proposed an efficient two-party protocol for training linear regression, logistic regression, and neural networks models. Rouhani et al. in DeepSecure \cite{deepsecure} proposed an approach for the scalable execution of deep learning models in the PPML setting that uses an optimized version of Yao's Garbled Circuit (GC)  protocol. Wagh et al. in SecureNN \cite{securenn} proposed a three-party computation protocol for privacy-preserving training and inference on convolutional neural networks.

In this paper, we consider PPML approaches using functional encryption. When machine learning is offered as a service, the model is stored on the server, and one or more clients are responsible for training. Sometimes there might be pre-trained models possessed by the server. In general, FHE-based machine learning models are trained on unencrypted data, and then their inference is obtained on encrypted data. In the FHE-based machine learning setting, a client sends encrypted data to the server; the server then performs tasks like classification using the pre-trained ML model on the data received from the client and produces the prediction results in the ciphertext. Even though the server performs computations on encrypted data, it doesn't learn anything and generates ciphertext predictions. So, only the owner of the data can see the actual result. However, in the case of functional encryption, the server generates the computation results in plaintext using a special key that enables partial decryption needed for computation (see Fig. \ref{fig:1}). The rest of the steps are identical to those of fully homomorphic encryption. In the realm of PPML, we found the idea of producing plaintext results over encrypted data without fully decrypting it to be intriguing for some of the applications. Note that both FHE and FE-based PPML works need high computation costs.

Consider a hospital that outsources the prediction of cardiovascular diseases to a machine learning as a service (MLaaS) company. Assume that this MLaaS company is responsible for performing machine learning operations on sensitive data like patient records. In such a scenario, a good ML model should predict the correct results and should also preserve the privacy of the patients' data. Earlier approaches to such cases can perform training of a machine learning model on the plain data and inference on homomorphically encrypted data. However, when an MLaaS is expected to produce the result in plain FHE-based approaches needs an additional key to decrypt the results predicted by the server. Functional encryption can be used to avoid sharing an additional key for the decryption of results inferred by the ML model.  

This SoK paper aims to study several research works in the field of functional encryption-based privacy-preserving machine learning. We investigate and examine the impact of functional encryption in the machine learning domain, i.e., machine learning computations on encrypted data. This is intended to give the research community a clear snapshot of FE-based PPML approaches. We only focus on FE-based PPML works out of many other existing PPML approaches.

Initiatives have been taken in recent years to develop FE-based PPML systems. FE-based PPML approaches are brodly categorized as 1) Inner-product FE-based (IPFE) approaches and 2) Quadratic FE-based (QFE) approaches. Ligier et al. \cite{ligier2017privacy} proposed an approach to perform privacy-preserving classification on data encrypted using IPFE. An IPFE-based deep neural network approach on MNIST dataset proposed by Xu et al. in \cite{xu2019cryptonn}. Panzade et al. in \cite{panzade2021towards} proposed a new approach to compute faster secure activation function based on function-hiding inner product encryption for PPML.   In \cite{ryffel2019partially}, Ryffel et al. introduced a system that used QFE with adversarial training to perform privacy-preserving predictions. Marc et al. \cite{marc2019privacy} presented the first fully-fledged FE libraries as well as a variety of applications for developing privacy-enhanced machine learning models.

The following are the primary contributions of this paper:
\begin{enumerate}
    \item We present a basic but substantial theoretical foundation to help researchers understand current approaches to FE-based PPML.
    \item We provide a thorough review of the literature on FE-based PPML, emphasizing the strengths and shortcomings of the various approaches to assessing how they supplement one another.
    \item We examine the current constraints that prohibit the implementation of existing FE-based PPML solutions in real-world settings, mostly due to issues with efficiency and usability. 
    \item We provide research directions to intensify existing works in terms of time performance and security that the research community may pursue in the coming years.
\end{enumerate}

\section{Background Knowledge}
\label{section:2}
Functional encryption is a generalization of public-key encryption that allows a key holder to compute a particular function of encrypted data using constrained secret keys \cite{boneh2011functional}. Here this function is called functionality. e.g., an FE scheme may be particularly designed to compute inner-products; in this case, the functionality becomes inner-product. In the FE scheme, a key management authority with a master secret key generates a secret key sk$_{fe}$; a decryptor can use that to compute a function on an encrypted message x. This section summarizes the two major functional encryption schemes, inner-product functional encryption, and quadratic functional encryption, used by privacy-preserving machine learning approaches. 

\subsection{Inner-product Functional Encryption}

The Decisional Diffie-Hellman assumption (DDH) underpins the method outlined by Abdalla et al. in \cite{abdalla2015simple}. Let GroupGenerator be a probabilistic polynomial-time (PPT) algorithm with input security parameter 1$^\lambda$, which produces a triplet ($\mathbb{G}$, $p$, $g$), where $\mathbb{G}$ is a group of order p created by g $in$ $\mathbb{G}$. The tuples (g, g$^a$, g$^b$, g$^{ab}$) and (g, g$^a$, g$^b$, g$^c$) are computationally indistinguishable, according to DDH, where ($\mathbb{G}$, $p$, $g$) $\leftarrow$ GroupGenerator(1$^\lambda$), and a, b, c $\in$ Z$_p$ are chosen uniformly and independently at random. 

The $\Pi_{\text {ipfe }}=$ ( Setup, Encrypt, KeyDerivation, Decrypt) FE scheme for IPFE in DDH is as follows: \newline
\textbf{Setup(1$^\lambda$, 1$^l$):} This algorithm samples ($\mathbb{G}$,$p$,$g$)$\leftarrow$ GroupGenerator(1$^\lambda$) and s = (s$_1$,......s$_l$)$\leftarrow$ Z$_{p}^{l}$, sets mpk = $\left(h_{i}=g^{s_{i}}\right)_{i \in[\ell]}$ and msk = s and finally returns a pair of (mpk, msk). \newline
\textbf{Encrypt(mpk, x):} This algorithm takes mpk and message x  = (x$_1$,......x$_l$) $\in$ Z$_{p}^{l}$ as input, chooses random number r $\leftarrow$ Z$_p$, computes Ct$_0$ = g$^r$ and, for each i $\in$ [l], Ct$_i$ = h$_i$$^r$. g$^{{x}_{i}}$ and returns ciphertext C$_t$. \newline
\textbf{KeyDerivation(msk, y):} This algorithm takes msk and vector y=(y$_1$,......y$_l$) $\in$ Z$_p$ as input and outputs key sk$_{fe}$. \newline
\textbf{Decrypt(mpk, Ct, sk$_{fe}$):} This algorithm takes the master public key, ciphertext and sk$_{fe}$ for vector y as input and outputs the discrete logarithm in basis g of \vspace{1em}

\hspace{7em}
$
\prod_{i \in [l]}^{} 
Ct_{i}^{y_{i}}/Ct_{0}^{sk_{fe}}
$
\vspace{1em} \\
\textbf{Correctness:}
The method's correctness is demonstrated as follows \cite{abdalla2015simple}: \newline
$\forall$ (mpk,msk) $\leftarrow$ Setup(1$^\lambda$, 1$^l$), all y $\in$ Z$_{p}^{l}$ and x $\in$ Z$_{p}^{l}$ \\ for sk$_{fe}$ $\leftarrow$ KeyDerivation(msk, y) and Ct $\leftarrow$ Encrypt(mpk,x).
\newline
Decrypt(mpk, $C_t, sk_{fe}$) \vspace{0.5em} \\= $ \frac{\prod_{i \in [l]}^{} 
Ct_{i}^{y_{i}}}{Ct_{0}^{sk_{fe}}} $ \vspace{0.5em}
\\$= \frac{\prod_{i \in [l]}^{} (g^{s_{i}r+x_{i})^{y_i}}}{ g^{r(\sum{i \in [l]}{}y_{i}s_{i})}      } 
 $\vspace{0.5em}
 \newline 
 $ = g^{\sum{i \in [l]}^{}y_{i}s_{i}r + \sum{i \in [l]}{}y_{i}x_{i} - r(\sum{i \in [l]}{}y_{i}s_{i})} $ \vspace{0.5em}
\\  = $g^{\sum{i \in [l]}^{}y_{i}x_{i}} $ \vspace{0.5em}
\\$= g^{<x, y>}$

\begin{table}
  \caption{Symbols and acronyms used in the paper}
  \begin{tabular}{ccl}
    \toprule
    Acronym / symbol & Description\\
    \midrule
    mpk & Master Public Key \\
    msk & Master Secret Key \\
    sk$_{fe}$ & FE Key for IPFE \\
$    \mathrm{sk}_{qe}$ & FE key for QFE \\    
  Ct & Ciphertext \\
  IND-CPA & security against chosen-plaintext attacks \\
  ERT & Extremely Randomized Trees \\
  IPFE & Inner-product Functional Encryption \\
  QFE & Quadratic Functional Encryption \\
    \bottomrule
  \end{tabular}
\end{table}
\subsection{Quadratic Functional Encryption}
Quadratic functional encryption scheme uses bilinear groups (also known as pairing groups), has been proposed by \cite{boneh2003identity, joux2004one}. In the case of QFE-based PPML, we refer to schemes proposed by Ryffel et al. in \cite{ryffel2019partially}. Here, GroupGenerator is a PPT algorithm on inputting $1^{\lambda}$ returns $\mathcal{P G} =\left(\mathbb{G}_{1}, \mathbb{G}_{2}, p, g_{1}, g_{2}, e\right)$ of an asymmetric bilinear group, where $\mathbb{G}_{1}$ and $\mathbb{G}_{2}$ are cyclic groups of prime order $p$ (for a $2 \lambda$-bit prime $p$ ) and $g_{1}$ and $g_{2}$ are generators of $\mathrm{G}_{1}$ and $\mathrm{G}_{2}$, respectively. The application $e: \mathbb{G}_{1} \times \mathbb{G}_{2} \rightarrow \mathbb{G}_{T}$ is an admissible pairing i.e. it can be efficiently computable, non-degenerated, and bilinear: $e\left(g_{1}^{\alpha}, g_{2}^{\beta}\right)=e\left(g_{1}, g_{2}\right)^{\alpha \beta}$ for any scalars $\alpha, \beta \in \mathbb{Z}_{p}$. Therefore, $g_{T}:=e\left(g_{1}, g_{2}\right)$ which makes the group $\mathbb{G}_{T}$ of order $p$, where $p$ is prime. For any $s \in\{1,2, T\}, n \in \mathbb{N},$ and vector $u:=\left(\begin{array}{c}u_{1} \\ \vdots \\ u_{n}\end{array}\right) \in \mathbb{Z}_{p}^{n},$ it is denoted by
$g_{s}^{u}:=\left(\begin{array}{c}g_{s}^{u_{1}} \\ \vdots \\ g_{s}^{u_{n}}\end{array}\right) \in \mathbb{G}_{s}^{n} .$ \\ Similarly, for any vectors $u \in \mathbb{Z}_{p}^{n}, v \in \mathbb{Z}_{p}^{n},$ It is denoted by $e\left(g_{1}^{u}, g_{2}^{v}\right)=$
$\prod_{1-1} e\left(g_{1}, g_{2}\right)^{\mathbf{u}_{t} \cdot \bar{v}_{1}}=e\left(g_{1}, g_{2}\right)^{\mathbf{u} \cdot \boldsymbol{v}} \in \mathbb{G}_{T},$ since $\boldsymbol{u} \cdot \boldsymbol{v}$ denotes the inner-product between the vectors $\boldsymbol{u}$ and
$v,$ that is: $\boldsymbol{u} \cdot \boldsymbol{v}:=\sum_{\mathrm{i}=1}^{n} u_{1} v_{i}$.\newline \newline
Ryffel et al. in \cite{ryffel2019partially} build an efficient FE scheme shown below for the set of functions defined, for all $n, B_{x}, B_{y}, B_{f} \in \mathbb{N}^{*},$ as $\mathcal{F}_{n, B_{x}, B_{y}, B_{f}}=\left\{f:\left[-B_{x}, B_{x}\right]^{n} \times\left[-B_{y}, B_{y}\right]^{n} \rightarrow \mathbb{Z}\right\}$ where the functions $f$ $\in \mathcal{F}_{n, B_{x}, B_{y}, B_{f}}$ are expressed as a set of bounded coefficients $\left\{f_{i, j} \in\left[-B_{f}, B_{f}\right]\right\}_{i, j \in[n]},$ and for all vectors $\boldsymbol{x} \in\left[-B_{x}, B_{x}\right]^{n}, \boldsymbol{y} \in\left[-B_{y}, B_{y}\right],$ :
\[
f(\boldsymbol{x}, \boldsymbol{y})=\sum_{i, j \in[n]} f_{i, j} x_{i} y_{j}
\]
FE scheme is explained as follows: \vspace{1em} \newline 
\textbf{Setup(1$^{\lambda}, \mathcal{F}_{n, B_{x}, B_{y}, B_{f}})$:} \newline
$\mathcal{PG}:=\left(\mathbb{G}_{1}, \mathbb{G}_{2}, p, g_{1}, g_{2}, e\right) \leftarrow \operatorname{GroupGenenerator}\left(1^{\lambda}\right),\\ \boldsymbol{s}, \boldsymbol{t} \leftarrow \mathbb{\mathbb { Z }}_{p}^{n}, \mathbf{m s k}:=(\boldsymbol{s}, \boldsymbol{t}), \\ \mathbf{m p k}:=\left(\mathcal{P} \mathcal{G}, g_{1}^{\boldsymbol{s}}, g_{2}^{t}\right)$\\
Return (mpk, msk).
\newline \newline \newline
\textbf{Encrypt(mpk,(x,y)):} \newline
$\gamma \leftarrow \mathbb{Z}_{p}, \mathbf{W} \leftarrow \mathrm{GL}_{2}, \text { for all } i \in[n], \\
\boldsymbol{a}_{i}:=\left(\mathbf{W}^{-1}\right)^{\top}\left(\begin{array}{c}
x_{i} \\
\gamma s_{i}
\end{array}\right), \boldsymbol{b}_{i}:=\mathbf{W}\left(\begin{array}{c}
y_{i} \\
-t_{i}
\end{array}\right)$
\\ \text {Return Ct}:=$\left(g_{1}^{\gamma},\left\{g_{1}^{a_{i}}, g_{2}^{b_{i}}\right\}_{i \in[n]}\right) \in \mathbb{G}_{1} \times\left(\mathbb{G}_{1}^{2} \times \mathbb{G}_{2}^{2}\right)^{n}$ \vspace{1em} \newline 
\textbf{KeyDerivation(msk,f):}  \vspace{1em} \newline
Return $\mathrm{sk}_{qe}:=\left(g_{2}^{f(s, t)}, f\right) \in \mathbb{G}_{2} \times \mathcal{F}_{n, B_{x}, B_{y}, B_{f}}$ \vspace{1em} \newline 
\textbf{Decrypt( mpk, Ct}\textbf{:=(g$_{1}^{\gamma},\{g_{1}^{a_{i}}, g_{2}^{b_{i}})\}_{i \in[n]}$}), \\
$\mathrm{sk}_{qe}:=(g_{2}^{f(s, t)}, f)\textbf{ ):} \vspace{1em}\\
out:=e\left(g_{1}^{\gamma}, g_{2}^{f(s, t)}\right) \cdot \prod_{i, j \in[n]} e\left(g_{1}^{a_{i}}, g_{2}^{b_{i}}\right)^{f_{i, j}}$ \vspace{1em}\\ 
\text{Return} $\log(out) \in \mathbb{Z}$ \vspace{1em}\newline
\textbf{Correctness:} \vspace{1em} \newline 
For all $i, j \in[n]$:
$$
e\left(g_{1}^{d_{i}}, g_{2}^{b_{j}}\right)=g_{T}^{d_{i} \cdot b_{j}}=g_{T}^{x_{i} y_{j}-\gamma s_{i} t_{j}}
$$
since
$$
\begin{aligned}
\vec{a}_{i} \cdot \vec{b}_{j} &=\left(\left(\mathbf{W}^{-1}\right)^{\top}\left(\begin{array}{c}
x_{i} \\
\gamma s_{i}
\end{array}\right)\right)^{\top} \cdot\left(\mathbf{W}\left(\begin{array}{c}
y_{j} \\
-t_{j}
\end{array}\right)\right) \\
&=\left(\begin{array}{c}
x_{i} \\
\gamma s_{i}
\end{array}\right)^{\top} \mathbf{W}^{-1} \mathbf{W}\left(\begin{array}{c}
y_{j} \\
-t_{j}
\end{array}\right)= x_{i} y_{j}-\gamma s_{i} t_{j}
\end{aligned}
$$
Therefore,
\newline
\[
 \text {out}=e\left(g_{1}^{\gamma}, g_{2}^{q(\vec{s}, \vec{t})}\right) \cdot \prod_{i, j} e\left(g_{1}^{\vec{a}_{i}}, g_{2}^{\vec{b}_{i}}\right)^{q_{i, j}}\]
  \hspace{5em}
$ =g_{T}^{\gamma q(\vec{s}, \vec{t})} \cdot g_{T}^{\sum_{i, j} q_{i, j} x_{i} y_{j}-\gamma q_{i, j} s_{i} t_{j}} $\\ 
 
 \hspace{3.2em}
 $=g_{T}^{\gamma q(\vec{s}, \vec{t})} \cdot g_{T}^{q(\vec{x}, \vec{y})-\gamma q(\vec{s}, \vec{t})}=g_{T}^{q(\vec{x}, \vec{y})} $

\vspace{0.5em}
We refer the readers to \cite{abdalla2015simple} and \cite{baltico2017practical, ryffel2019partially} for more cryptographic details on IPFE and QFE schemes, respectively.
\subsection{Neural networks} The artificial neural network, often known as a neural network, is a machine learning model that is hierarchical and non-linear, with several layers and several neurons in each layer. Each layer of a neural network processes the input provided by the previous layer before passing it on to the next.
\begin{itemize}
    \item \textbf{Input layer:}  The preprocessed raw data or features extracted from raw data in a particular format make up the first layer of the neural network.
    \item \textbf{Hidden Layer:}  A neural network can have one or more hidden layers. The First Hidden layer's neurons are linked to the input layer and followed by an activation function. Further hidden layers are fed with the previous layer's output. Weight values are associated with layers, and they are updated during the forward and backpropagation processes until convergence.
    \item \textbf{Activation function:}  The activation function of a neuron in a neural network determines the output of that neuron given a single or group of inputs. In machine learning, there are several activation functions such as sigmoid, Rectified Linear Unit (ReLU), and tanh. The ReLU activation function is an example of one of the most frequent activation functions. If the input value is less than zero, the ReLU activation function returns zero, and if it is larger than zero, it returns the same input.
    \item \textbf{Output layer:} The output layer of a neural network is the final layer of neurons that provides the network's output.
\end{itemize}

\subsection{Polynomial Neural Network}
\label{subsection:2.4}
Polynomial neural networks are the kinds of neural networks that primarily facilitate linear components like fully connected layers, convolutions with average pooling, and activation functions are approximated using polynomials. They have demonstrated fairly high accuracy for the relatively simple benchmarks in image recognition tasks \cite{cryptonet, badawi2018alexnet}. They have also been used to propose the applicability of novel machine learning protocols in various early-stage implementations, as presented in \cite{bourse2018fast, chillotti2016faster}. The simplicity of the operations upon which polynomial neural networks are built ensures high efficiency, particularly for gradient computations. Research works such as \cite{livni2014computational} have demonstrated that they can reach convergence rates comparable to those of networks with non-linear activation functions. 

\section{Scope and Methodology}

Our analysis of the FE-based PPML methodologies is divided into two parts. To begin, we conduct a thorough review of existing methodologies and emphasize their salient features and functions. Second, we evaluate these tools in practice by comparing their usability, complexity, and performance across various case study scenarios. We integrate quantitative performance analysis with a qualitative review, highlighting the difficulties inherent in designing various applications.
The secure computation ecosystem encompasses a wide variety of tools. On the low level, there exist math libraries that facilitate the construction of FE implementations, for example, by efficiently implementing algorithms useful for generic lattice-based cryptography. Then there are FE libraries that implement certain schemes and provide slightly more advanced APIs, such as setup, encrypt, key generation, and decrypt. These libraries abstract away computation features like parameter selection, encryption, and decryption by providing a higher-level language in which developers can implement their computation.

The primary objective of this work is to comprehend the landscape of MLaaS in data-sensitive scenarios via privacy-preserving machine learning computing, for example, when data supplied to third parties for processing is encrypted. PPML ensures the privacy and confidentiality of input data. Additionally, they alleviate excessive pressures on the client endpoint in computing as cloud server does most of the computing part. Finally, they may be used in machine learning situations where clients can contribute data toward a training or the inference goal. PPML is frequently used in conjunction with other approaches. We discuss the application of all PPML approaches based on FE just at these crossing places.  Although fully homomorphic encryption is a widely used technique in PPML, we only include FE-based approaches because of fewer available articles to the research community on this topic. Also, this work does not cover federated learning, differential privacy, and secure multi-party computation due to their different computational infrastructure requirements and implementation. 

PPML that is based on FE has efficiency and usability concerns. FE-based PPML is limited in its computing capabilities due to performance concerns associated with extensive computations. Even though they have been frequently used in other contexts, like private data aggregation and statistics, their application to machine learning is not simple. Moreover, the works focusing on FE-based PPML are significantly lower in number. So, we consider a solution efficient if it offers enhancements to previous methods that reduce their runtime on machine learning applications. 

The second problem is associated with the techniques' deployability. Numerous frameworks and technologies make advanced machine learning accessible to data scientists who are not necessarily professionals in computer sciences. However, adopting these frameworks for use with PPML is challenging. We consider usability enhancements in this area if the proposal fits one of the following two requirements. To start, if it makes the solution more adaptable to existing machine learning frameworks (for example,by giving tools that reduce total programming effort). As such, our study incorporates works that propose Application Programming Interfaces (APIs), compilers, or other significant practical tools that aid in the implementation and deployment of theoretical ideas into real applications, hence increasing their usability. Second, when an open-source implementation supplements the concept.
Apart from facilitating future revisions to the approach, open-source implementations ensure that the results are reproducible. Additionally, we verify whether the work contains references to open-source implementations or is released independently (for example, by visiting the authors' websites or GitHub repositories). 

Furthermore, we investigate and analyze two parameters of such open-source repositories:  i) the code's adherence with the theoretical claims (for example, if it employs the security methods and features stated in the paper); and ii) how well the source code is maintained or integrated with other current frameworks (for example, the version updates since it was released). 

In summary, we examine the following for each of the proposals utilizing FE-based PPML: (i) the problem being addressed (training, inference, or both), (ii) the machine learning model used, (iii) the specific FE techniques involved (IPFE or QFE), and (iv) the efficiency and usability considerations examined.

We execute searches in multiple scholarly repositories and databases, mainly Google scholar, IEEE Xplore and ACM digital library  for relevant publications using specified keywords such as privacy-preserving machine learning, functional encryption, secure computation, and computation over encrypted data. Secondly, we select those focused on FE-based PPML and have been published in renowned venues like NeurIPS, PETS, IEEE S\&P, CRYPTO, EUROCRYPT, TCC, CCS. We then read their abstracts and methodologies to determine if they fall within the topic of our research. This provides us with an initial collection of works, which we meticulously studied. Finally, we use snowball sampling to add additional suitable works by using references from the first collection of papers. We made every effort to include all significant works implementing PPML using FE.

\section{Overview of FE-based PPML models and libraries}

\subsection{What has been done?}
There are two variations of the FE-based PPML methodologies available in the literature. The first one uses inner-product functional encryption, whereas the other one uses quadratic functional encryption. The IPFE-based methodologies involve training and inference, whereas the QFE methodologies involve simply the inference stage of machine learning.

\subsubsection{Inner-product FE-based Machine Learning}
In this type of methodology, the inputs are encrypted using inner-product functional encryption. Then during the activation, inner products between encrypted inputs and weight matrices are unfolded based on the special property of FE. Later, the neural network operations are done similarly to regular neural networks. Here both forward propagation and backpropagation can be made secure using FE. This methodology supports both training and prediction over encrypted data, unlike the quadratic FE-based approaches.
\subsubsection{Quadratic FE-based Machine Learning}
In quadratic FE-based methodologies, the training phase happens similar to the regular neural networks. It is also worthwhile to note that they train the neural networks with plain data. In the prediction phase, encrypted data is fed to the neural network. It undergoes the process of polynomial approximation, and then the other steps are applied similar to regular neural networks. These approaches are found to be faster than inner-product FE-based approaches. 
\subsection{Cryptography libraries for Functional Encryption implementation}
Presently, there are only two dedicated libraries that focus on implementing the state-of-the-art FE schemes. The first one is called CiFEr, and the other is GoFE.  Three entities are involved in functional encryption and decryption: an encryptor, a decryptor, and a key management authority. An encryptor encrypts the data and obtains ciphertext. The decryptor decrypts the ciphertext received from the encryptor. The key management authority handles the responsibility of generating a variety of cryptographic keys. In the FE scheme, based on the involvement of encryptors, it can be either single input or multi-input. We detailed both of these libraries and supporting cryptography libraries used in the FE-based PPML works as follows:
\subsubsection{CiFEr}
CiFEr \cite{marc2019privacy} is a functional encryption library developed by the FENTEC group that is written in the C language. This library provides developers to use functions to perform various FE operations like encrypt, decrypt, and key generation. These libraries are built in such a way that the predefined functions can be directly called without setting many parameters. Here, the cryptographic key generation is abstract to the user, and users set the security parameters in terms of bits. It provides both single-input and multi-input FE implementations commonly observed in FE-based PPML approaches. 
\subsubsection{GoFE}
GoFE \cite{marc2019privacy} is another library provided by the FENTEC group that is written in Golang. Like CiFEr, it also provides an option to use FE functions. Both of CiFEr and GoFE have the same set of state-of-the-art functional encryption implementations. Their performance is based on the underlying programming languages.
\subsubsection{FLINT}
FLINT \cite{hart2013flint} is a cryptography library that is used for performing number theory-related operations. It is written in 
C. Unlike GoFE and CiFEr, it does not dedicatedly provide FE functionality. In some of the FE-based works, this library is used for the implementation of FE schemes.
\subsubsection{Charm}
Charm \cite{akinyele2013charm} is a framework designed for implementing various advanced cryptosystems. It is built using Python to decrease development time and code complexity while fostering component reuse.

\subsubsection{PBC}
Pairing-based Cryptography (PBC) \cite{lynn2006pbc} is a C library that enables rapid development of cryptosystems based on pairings. It implements a bilinear cyclic group abstractly, hiding the programmer from mathematical details. Both PBC and Charm are used in some of the FE-based PPML implementations. 
\begin{table*}[!ht]
\centering
\begin{threeparttable}[b]
\caption{\centering Comparison of FE-based PPML models in terms of functionality}
\begin{tabular}{ccccccc}
    \toprule
    Research works & FE type & Training\tnote{1} & Prediction\tnote{2} &  ML model & Security \\
    \midrule
    Ligier et al. \cite{ligier2017privacy} & IPFE\tnote{3} & \priority{0} & \priority{100} & ERT\tnote{4} & Selective IND-CPA\tnote{5} \\
    Xu et al. \cite{xu2019cryptonn} & IPFE & \priority{100} & \priority{100} & 5 layer NN\tnote{6} & Selective IND-CPA\\
    Sans et al. \cite{sans2018reading} & QFE\tnote{7} & \priority{0} &\priority{100} & 2 layer NN & Adaptive IND-CPA\tnote{8}\\
    Ryffel et al. \cite{ryffel2019partially} & QFE & \priority{0} &\priority{100} & 2 layer NN & Adaptive IND-CPA \\
    Carpov et al. \cite{carpov2020illuminating} & IPFE and QFE & \priority{0}&\priority{100} & 2 layer NN & Adaptive IND-CPA\\
    \bottomrule
\end{tabular}

\begin{tablenotes}
\item [1] Training on encrypted data
\item [2] Prediction on encrypted data
\item [3] Inner-product Functional Encryption
\item [4] Extremely Randomized Trees
\item [5] Selective security against chosen-plaintext attack
\item [6] Neural Network
\item [7] Quadratic Functional Encryption
\item [8] Adaptive security against chosen-plaintext attack
\end{tablenotes}
\end{threeparttable}
\end{table*}
\section{Insights}

Recall that we consider the FE-based PPML works focused on training and/or inference over encrypted data. 

\subsection{Threat model}
\textbf{Assumption:}
In FE-based PPML, it is assumed that there is a trusted independent key management authority that is responsible for the generation of the required keys. Such assumptions are common in all functional encryption schemes. 

FE-based PPML techniques adhere to an honest but curious security model, in which both parties comply with the protocol while attempting to gain as much information as possible. In this case, the server follows the protocol but tries to learn additional information. Typically, the approach includes three primary components: a key management authority, a server, and a client. The key management authority generates encryption and decryption keys. FE requires three distinct keys, namely master public key (mpk), master secret key (msk), and functional encryption key(sk$_{fe}$), that operate slightly differently than public-key cryptography techniques as discussed in section \ref{section:2}.

The research works in this field assume an MLaaS setting. The server holds the trained model, and the client provides the data to the server and receives the prediction results using the learned model. Additionally, in some circumstances, the capability of training models over encrypted data may be available.

The client firstly encrypts the data using the master public key before sending it to the server. In some scenarios, the client may require to preprocess the data before encryption; this varies by application. For example, computer vision applications may require scaling or transforming images prior to encryption. Normalization and standardization may be necessary in the case of structured data.

The server holds the model. The model is developed by training the neural network using client-supplied data. Because the server cannot see the data sent by the client, it must obtain the functional encryption secret key sk$_{fe}$ in order to execute functionality-based computations on the data. Either polynomial approximation uses such computations for QFE-based methods or inner-product computation in IPFE-based methods discussed in the later sections.
\begin{figure}[!ht]
    \centering
    \includegraphics[width=\linewidth]{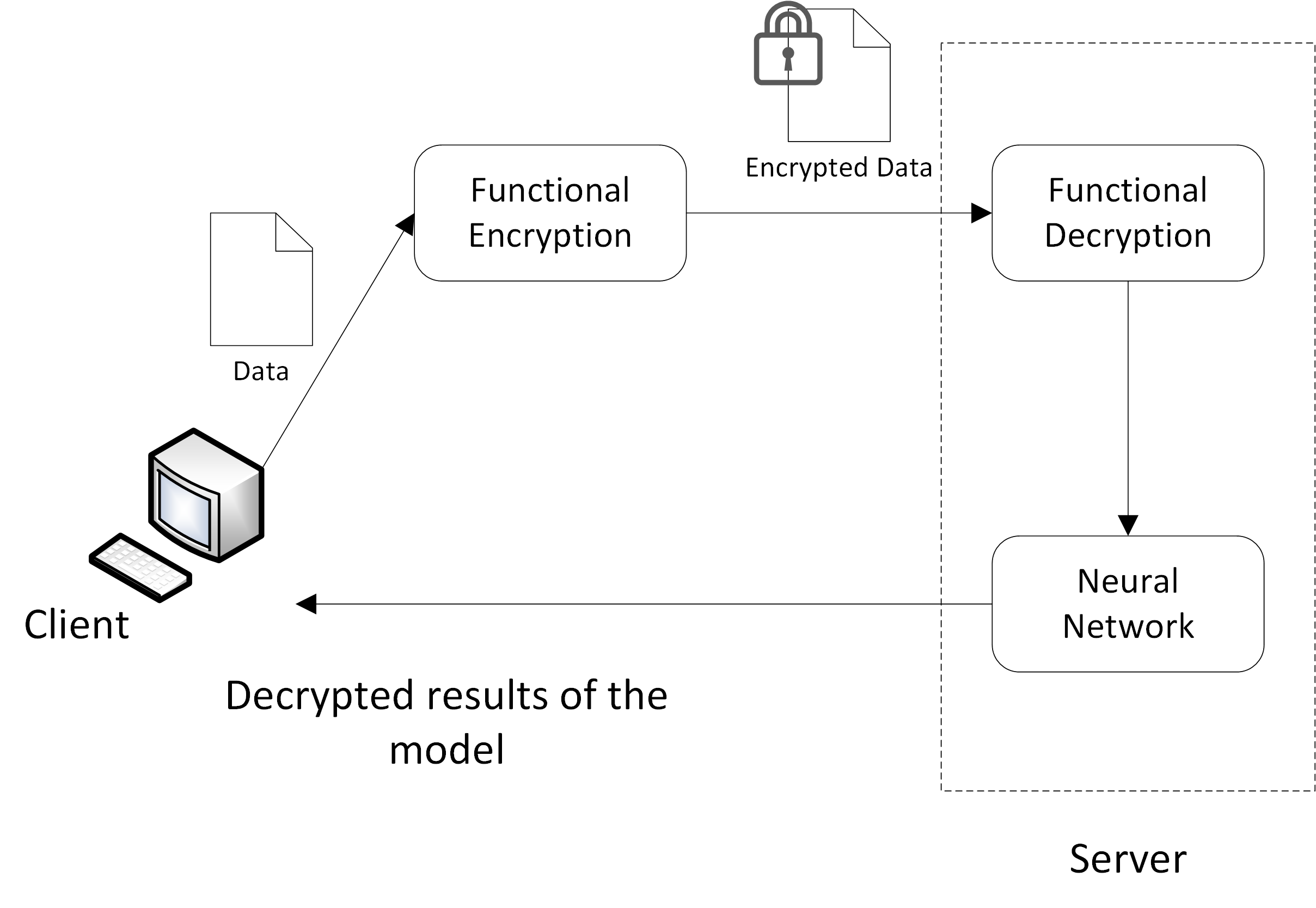}
    \caption{\centering IPFE-based PPML}
    \label{fig:2}
\end{figure}
\subsection{IPFE-based PPML}
Fig. \ref{fig:2} depicts IPFE-based PPML. Out of the available works in this field dedicated to FE-based PPML \cite{ligier2017privacy} and \cite{xu2019cryptonn} use inner-product FE schemes. 

\subsubsection{Functional encryption scheme}
The work proposed in \cite{ligier2017privacy} and \cite{xu2019cryptonn} use the inner product schemes proposed in \cite{agrawal2016fully} and \cite{abdalla2015simple}, respectively. 

\subsubsection{Training phase}
In the approach proposed by \cite{ligier2017privacy}, Extremely Randomized Trees are used as a machine learning model. These trees are nothing but an ensemble learning model that uses decision trees. In their work, they perform the training on plain data that is similar to regular machine learning models. For the inference, they used IPFE. 

The methodology proposed in [48] uses a LeNet like 5-layer neural network. In their work based on IPFE, inputs are first encrypted by the client and then sent to a server. The server obtains the FE key and runs the IPFE decrypt function in order to compute the activation results of the first hidden layer. 
The computation between the input vector and weight matrix is done as follows:  \newline 

\hspace{2em}
 A = ReLU(sk$_{fe}$(W) * Encrypt(X) + b) \vspace{1em} \\ 
This output obtained after the first hidden layer is then fed through the remaining layers of the neural network. So it can be said that these methodologies can work entirely on FE encrypted data.
The training method over encrypted data proposed by \cite{xu2019cryptonn} requires 57 hours to train the 5-layer neural network model. Regular neural network with the same settings on non-encrypted data takes 4 hours. This shows a massive gap between computation on encrypted data over non-encrypted data considering neural networks.

\subsubsection{Inference phase}
The methodology proposed in \cite{ligier2017privacy} can perform predictions on encryption data with a prediction time of fewer than 0.1 seconds. Also, it produces a validation accuracy of 95.64 \%. The work proposed in \cite{xu2019cryptonn} can run the inference on encrypted data by following the similar use of IPFE decryption functions for the inference. Their model achieves a validation accuracy of  95.49 \%.

\subsection{QFE-based PPML}
\begin{figure}[!ht]
    \centering
    \includegraphics[width=\linewidth]{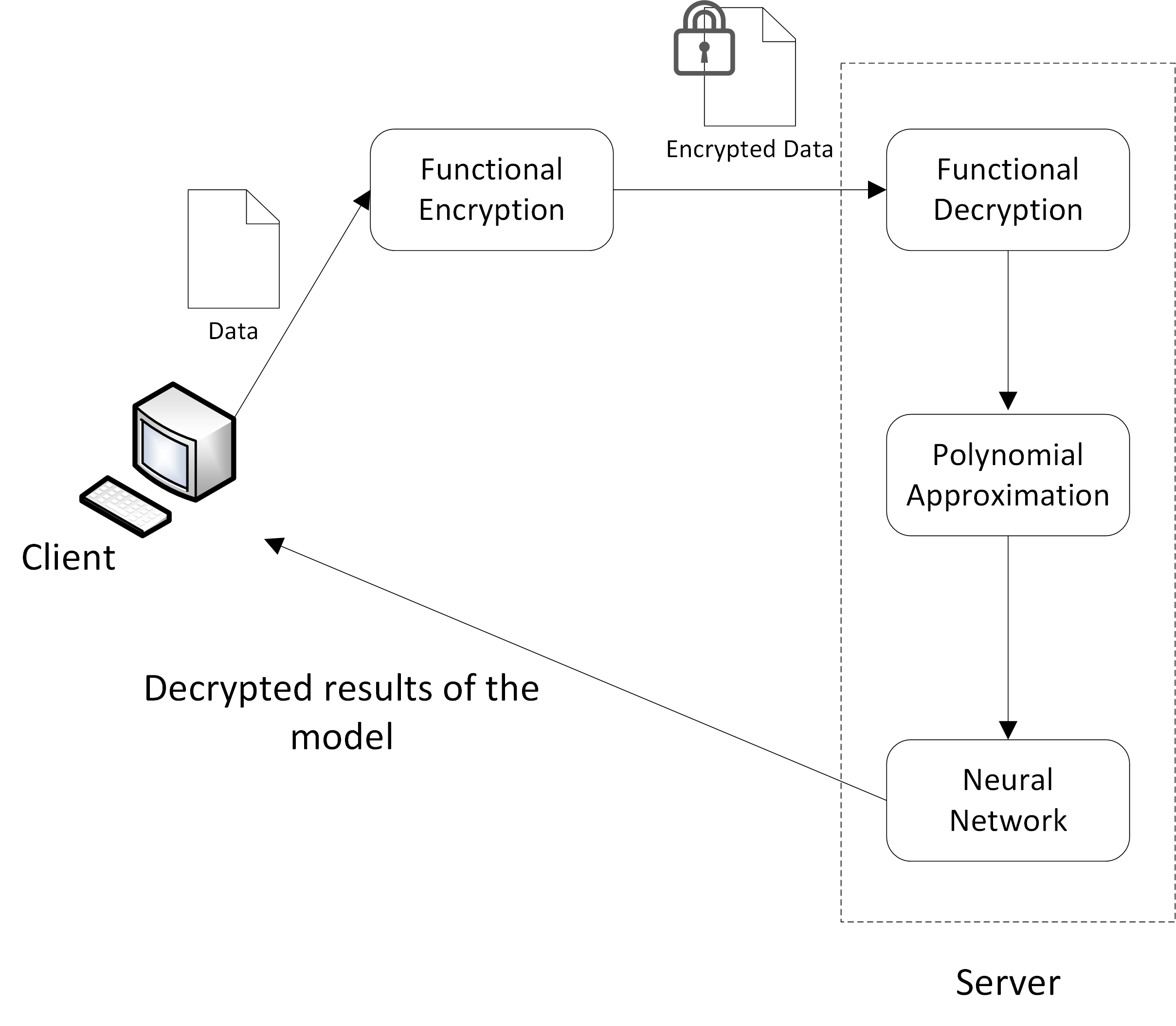}
    \caption{\centering QFE-based PPML}
    \label{fig:3}
\end{figure}

Fig. \ref{fig:3} depicts QFE-based PPML.
Three works use quadratic functional encryption schemes, which are defined and shown utilization towards machine learning problems in \cite{sans2018reading,ryffel2019partially, carpov2020illuminating}. 
\subsubsection{Functional encryption scheme}
This type of research works either use the similar or modified approach proposed by Baltico et al. \cite{baltico2017practical}. Works proposed by \cite{sans2018reading} and \cite{ryffel2019partially} are related to each other in terms of the FE scheme and neural networks used. They proposed a novel functional encryption scheme that can be used on multivariate quadratic polynomials.
\begin{table*}[!ht]
\centering
\caption{\centering Comparison of FE-based PPML models in terms of performance}
\begin{tabular}{cccccccc}
    \toprule
    Research works & Year& Training time  & Prediction time & FE key generation & Encryption time & Decryption & Accuracy \\
    \midrule
    Ligier et al. \cite{ligier2017privacy} &2017 & NA & $\leq$ 0.1s & 12 ms & 150 ms & 69s & 95.64\%\\
    Xu et al. \cite{xu2019cryptonn} & 2018 & 57 hrs & not specified & not specified & not specified & not specified & 95.49\% \\
    Sans et al. \cite{sans2018reading} & 2019 & NA & a few seconds & 8ms & 8.1s & 3.3s & 97.54\% \\
    Ryffel et al. \cite{ryffel2019partially} & 2019 & NA & $\leq$ 3 ms & 94 $\pm$ 5ms & 12.1 $\pm$ 0.3s & 24 $\pm$ 9ms & 97.7\%\\
    Carpov et al. \cite{carpov2020illuminating} & 2020 & NA & not specified & not specified & not specified & not specified & 90 \% \\
    \bottomrule
\end{tabular}
\end{table*}

\begin{table*} [!ht]
\centering
\begin{threeparttable}
\caption{\centering Implementation details}
\label{table:4}
\begin{tabular}{ccccccc}
    \toprule
    Research works & Dataset & FE Libraries & Python Libraries & System specifications \\
    \midrule
    Ligier et al. \cite{ligier2017privacy} & MNIST\tnote{1} & FLINT\tnote{2} & sklearn \cite{pedregosa2011scikit} & Intel Core i7-4650U, 8GB RAM\\
    Xu et al. \cite{xu2019cryptonn} & MNIST & Charm\tnote{3} & Numpy \cite{van2011numpy} & Intel Core i7, 8GB RAM \\
    Sans et al. \cite{sans2018reading} & MNIST & PBC\tnote{4} and Charm & Tensorflow \cite{abadi2015tensorflow} & Intel Core i5-6440HQ, 8GB RAM \\
    Ryffel et al. \cite{ryffel2019partially} & MNIST & PBC and Charm & Tensorflow & Intel Core i7, 16GB RAM\\
    Carpov et al. \cite{carpov2020illuminating} & MNIST, Census Income & not specified & Keras \cite{chollet2018keras} & not specified  \\
    \bottomrule
\end{tabular}
\begin{tablenotes}
\item [1] MNIST handwritten digits dataset \cite{lecun2010mnist}
\item [2] Fast Library For Number Theory \cite{hart2013flint}
\item [3] A framework for rapidly prototyping cryptosystems \cite{akinyele2013charm}
\item [4] Pairing-based Cryptograph library \cite{lynn2006pbc}
\item [5] 1994 Census database \cite{kohavi1996scaling}
\end{tablenotes}
\end{threeparttable}
\end{table*}

\subsubsection{Training phase}
The training phase in this type of methodology is done on plain data similar to regular neural networks \cite{sans2018reading} and \cite{carpov2020illuminating}. In work proposed by \cite{ryffel2019partially}, the adversarial training approach is used in order to avoid information leakage after functional decryption. They optimized both the primary classification objective and the opposite of the collateral objective of a particular simulated adversary simultaneously using adversarial training. 
\subsubsection{Inference phase}

In the inference phase of all the works proposed in \cite{sans2018reading, ryffel2019partially, carpov2020illuminating}, polynomial neural networks are used. Here they use two-layer neural networks. Wherever quadratic functional encryption is used, a special type of neural network called the polynomial neural network is used discussed in section \ref{subsection:2.4}. The approaches based on QFE have not yet been used for training the model over encrypted data because of the complexities involved in the cryptographical aspects. Also, the present functional encryption is defined for the usage for degree 2 polynomials. The methodology proposed by \cite{sans2018reading} and \cite{ryffel2019partially} requires fewer than 3 seconds to produce a prediction result. Moreover, in the case of QFE-based works, accuracy is shown a little higher. \cite{sans2018reading} and \cite{ryffel2019partially} obtain the accuracy of 97.54\% and 97.7\%, respectively.

\subsection{Security} 
\textbf{Theorem 1} \quad Under the decisional diffie hellman (DDH) assumption, IPFE scheme given in section 2.1 is \textit{selectively secure against chosen-plaintext attacks} (IND-IPFE-CPA). \\ \\
\textbf{Theorem 2} \quad Under the matrix decisional diffie hellman (MDDH) assumption, QFE scheme given in section 2.2 is \textit{adaptively secure against chosen-plaintext attacks} (IND-QFE-CPA).\\

We refer the readers to \cite{abdalla2015simple} and \cite{baltico2017practical, ryffel2019partially} for the security proofs of the theorems.

The security of the FE-based PPML approaches is dependent on the cryptographic security of the underlying FE schemes. They offer two types of security: adaptive security and selective security against chosen-plaintext attacks. If the security guarantee happens only for the messages known ahead of time, it is known as selective security. However, adaptive security is a concept where a security guarantee happens even for randomly chosen messages at any point in time. The approaches proposed in \cite{ligier2017privacy} and \cite{xu2019cryptonn} are selectively secure whereas the approaches proposed in \cite{sans2018reading}, \cite{ryffel2019partially} and \cite{modic2019privacy} show adaptive security based on the underlying crypto schemes.

\subsection{Information leakage and attacks} 
Although functional encryption schemes provide cryptographic security to the machine learning models, there is a risk of information leakage. The outputs obtained in both the IPFE and QFE based models are in plain after the FE decryptions are done. So, if the server tries to retrieve part of the information from the obtained plain data after FE computations, there is a possibility of information leakage. Carpov et al. in \cite{carpov2020illuminating} discusses the type of information leakage. Also, an adversarial training-based approach proposed in \cite{ryffel2019partially} can avoid information leakage to some extent. We believe there is a scope of attacks like model inversion attacks \cite{fredrikson2015model} and direct inference attacks that machine learning security researchers can study. In order to make the FE models deployable, this potential threat should be taken into consideration. 
\subsection{Implementation Details}
Almost all of the methodologies use MNIST dataset \cite{lecun2010mnist} for the experiments. MNIST contains 70,000 grayscale images of size 28x28 pixels of handwritten single digits from 0 to 9. Out of 70K images, the training set consists of 60K images, and the test set consists of 10K images. In \cite{carpov2020illuminating}, Census Income Dataset \cite{kohavi1996scaling} is used in addition to MNIST dataset. Census Income data was extracted from the 1994 Census bureau database. It is a multivariate dataset with 48842 instances. There are 14 attributes like age, work class, education with categorical and integer data. The task here is to predict whether a person's income exceeds  \$50K/yr. 

\textit{FE Libraries.} Ligier et al. in \cite{ligier2017privacy} implemented IPFE scheme using FLINT library \cite{hart2013flint} whereas \cite{xu2019cryptonn} used Charm libary \cite{akinyele2013charm} for implementation of IPFE scheme. Sans et al. and Ryffel et al. \cite{sans2018reading}, \cite{ryffel2019partially} used PBC (pairing-based cryptography) \cite{lynn2006pbc} and \cite{akinyele2013charm} library for implementation of QFE scheme. The work proposed in \cite{modic2019privacy} used GoFE libary proposed for the implementation of QFE scheme. We have already detailed these libraries in section 5.2. 

\textit{Python Libraries.} \cite{ligier2017privacy} used sklearn \cite{pedregosa2011scikit} in python for implementation of ML model. \cite{sans2018reading} and \cite{ryffel2019partially} implemented the ML model in Tensorflow \cite{abadi2015tensorflow} in python. \cite{xu2019cryptonn} uses just a Numpy \cite{van2011numpy} for the implementation of ML model whereas \cite{carpov2020illuminating} used Keras \cite{chollet2018keras} in python for the implementation of ML model. 

\textit{System specifications.}
Table \ref{table:4} summarizes the machine specifications used for performing experiments by the various authors. The experiments done by all the research works are on intel CPUs. As CPUs are involved, there is an issue of slow computation involved during training and prediction. This is far away from the faster computations done by today's GPU-based machine learning models.

\section{Discussion}

\subsection{Pros and Cons}
\textit{\quad To ensure the fully privacy-preserving machine learning system, both training and prediction phases over encrypted data should be integrated into the system.} 

Our studies observed that for a perfect PPML system, both training and prediction should be undergone over encrypted data. By doing so, the system ensures the privacy of the user's data for both tasks. As suggested in the example of privacy-preserving cardiovascular disease prediction service in section \ref{section:1}, such data of the user is highly confidential. In such cases, it would be great if the cloud server can train the possessed model without seeing the user's private data. Similarly, if predictions are to be done, it would also use the encrypted inputs. So, this interesting concept of training and prediction without seeing helps us develop a fully privacy-preserving machine learning system. Presently, only \cite{xu2019cryptonn} focuses on both of these tasks, but it is not efficient in terms of time. 

\textit{Considering information leakage and adaptive security simultaneously is highly important.}

Security of the FE-based PPML schemes is entirely dependant on the underlying FE scheme. IPFE is protected by selective security, whereas QFE is protected by selective as well as adaptive security against chosen-plaintext attack. There is a theoretical advancement by \cite{cryptoeprint:2020:209} towards making IPFE adaptively secured, but FE-based PPML works do not yet implement it. Also, information leakage proposed by \cite{carpov2020illuminating} is the only work that discusses this concept. This work also follows the traditional selective IPFE and adaptive QFE schemes. This raises a high need to consider both information leakage and adaptive security while building an FE-based PPML scheme. 

\textit{There is no fit-for-all methodology.}

None of the existing solutions considers all the criteria for a perfect PPML system. However, we can rank the methodologies from each group i.e one from the IPFE-based scheme and another from the QFE-based scheme. In the IPFE-based scheme, \cite{xu2019cryptonn} satisfies most of the criteria, whereas \cite{ryffel2019partially} performs the best in QFE-based methodologies. If prediction accuracy and privacy is considered, then \cite{ryffel2019partially} leads, whereas the overall performance of the system in terms of security, privacy, and accuracy is concerned, then \cite{xu2019cryptonn} leads over others.

\textit{FE Vs FHE.}

Both FHE and FE cryptosystems are based on post-quantum lattice-based cryptography. Apart from the key distinction, i.e., the additional key requirement of FHE, FE has some pros and cons. The FE can be the best choice when the computation results on the encrypted data are expected to be plain. Although this is an advantage, present FE schemes support only inner-product and quadratic computation on encrypted data. In this regard, FHE wins over FE as it gives computation capability for more complex polynomials. Both cryptosystems have the overhead of time and space complexity required for the encryption, decryption, and key generation processes. FHE is more developed in engineering aspects as leading companies like Microsoft \cite{chen2017simple} and IBM \cite{halevi2014algorithms} have been working on it. FE has not yet received attention and lacks engineering aspects. Due to this, even though FE and FHE have a set of complex operations, FHE uses available modern hardware accelerator devices and engineering solutions and wins over FE.

\subsection{Challenges and future research directions}

\subsubsection{Functionality enhancement in FE scheme}
The current research done on FE is limited to two functionality on integers: inner product and bilinear maps. Recall that functionality is a function computed over encrypted data to obtain decrypted results. As discussed, inner product functionality is used in the IPFE scheme, whereas bilinear maps are used in the QFE scheme. Because of this limited availability of functionalities, FE-based PPML methodologies still require enhancement. For example, QFE-based machine learning methods can use only 2-degree polynomial networks. Also, FE cannot perform Min/max and comparison operations. If enhancement in FE functionality is done, it will be helpful for PPML researchers to come up with practical solutions for supervised and unsupervised machine learning problems. However, these systems are not at a level to be used in real-world applications. 

\subsubsection{Improving efficiency}
Though FE-based PPML solutions have shown promising results on partially encrypted machine learning, there is a need to improve the efficiency. Ciphertexts and keys generated with larger security parameters are slower compared to other cryptographic approaches. This becomes a threat when computations are performed on large datasets. Therefore this efficiency stands as a big challenge in FE-based PPML methodologies.

\subsubsection{Improvement in Security and Privacy of FE scheme}
As discussed in the previous subsection, the FE-based PPML methodologies' security depends on the underlying FE scheme. IPFE schemes \cite{abdalla2015simple, agrawal2016fully} are selectively secured against chosen-plaintext attacks under the decisional-diffie hellman (DDH) assumption. QFE schemes \cite{ryffel2019partially, baltico2017practical} are adaptively secured against chosen plaintext attacks under Matrix Decisional Diffie-Hellman assumption. These FE schemes will lag in stricter security applications such as the defense domain. Information leakage in PPML based FE schemes needs to be explored and improved. In addition to this, machine learning attacks are also interesting to consider in this area. 

\subsubsection{Enhancing Privacy-preserving Neural Networks}
Today's FE-based works have shown their demonstration only up to 5-layer neural network. There is a scope to enhance the structure of neural networks. As discussed above, QFE-based methods are limited to using only degree-2 polynomial networks. So, the enhancement of these networks is somehow dependent on the underlying FE scheme. Implementation of various activation functions is also dependent on the underlying FE scheme. Apart from this, complex deep learning convolutional neural networks like VGGNet, AlexNet, and GoogleNet are far from implementing FE. Also, the methods like transfer learning can be applied if there exists a functionality that can decrypt the encrypted model and retrieve the saved parameters like general CNNs. 

\subsubsection{Training-centric PPML systems}
As the PPML is growing day by day, training over encrypted data is gaining the attention of researchers \cite{securenn}. Presently, FE-based PPML schemes proposed in the literature are more inference-centric. Although the work proposed in \cite{xu2019cryptonn} has successfully proposed a way for training a neural network using functional encryption, it lacks computational efficiency. There is a high need for PPML systems that can perform both training and inference over encrypted data.

\subsubsection{Training from multiple data sources}
As far as machine learning is concerned, there are methodologies that focus on training models from multiple data sources in the literature \cite{securenn}. The enhanced versions of the FE scheme, like multi-input functional encryption \cite{abdalla2017multi, brakerski2018multi}, can be used to serve this purpose. These schemes facilitate using multiple vectors to perform computation based on inner-product functionality. Such schemes could leverage the problem of training the model possessed by a cloud server for training from multiple data sources.

\subsubsection{Using Multi-authority and Decentralized extensions of FE}
The FE scheme used in PPML services requires a trusted third party or trusted authority to be involved in generating the keys. The scheme proposed by \cite{ambrona2021controlled} can be used to avoid the involvement of trusted authorities. In addition to this, there is a great scope for making the FE schemes used in PPML decentralized by using the approaches proposed by \cite{chotard2018decentralized} and \cite{chotard2020dynamic}.

\subsubsection{Need for open source library support}

Functional encryption has gained widespread attention from researchers as far as the theoretical aspects are concerned. However, it still lacks practical library implementations compared to its predecessor, fully homomorphic encryption. For example, HElib \cite{halevi2014algorithms} by IBM research group, and SEAL \cite{chen2017simple} by Microsoft research group are great tools for performing the fully homomorphic encryption-related task. Presently, commendable efforts have been taken by FENTEC group \cite{modic2019privacy} for providing C and Go language versions of FE libraries, namely CiFEr and GoFE. These libraries are in the development stage and are not yet ready to be deployed in production. So, there is a considerable need for practically applicable functional encryption libraries.

\subsubsection{Need for hardware acceleration support}
In all the implementations of FE-based PPML methodologies studied, computations are done on the CPU. Until now, there has been no GPU support available for FE implementation. Because of these limitations, we lack the opportunity to utilize the highest GPU compute available today. Real-world applications in machine learning are trained on massive data. So, performing computations in such applications is very time-consuming if done on the CPU. The work proposed by \cite{fpga} uses SoC implementation to provide acceleration support, but it needs special hardware. 
If easy-to-use hardware acceleration support using widespread GPU or FPGA is provided in such applications, it will speed up the computations faster. Such acceleration will eventually help in making FE deployable in real-world applications.

\subsubsection{Improving scalability and application to non-image datasets}

Present works focused on FE-based PPML are limited to using MNIST like small datasets for their experiments, and there are no implementations for larger datasets. In addition to this, the applications to non-image datasets like text corpus may also be explored. Therefore, this stands as a future research direction in improving FE utilization in real-world applications.

\section{Conclusion}
Privacy-preserving machine learning has gained widespread attention among industry and academic researchers in recent years. 
Although approaches based on fully homomorphic encryption, secure multi-party computation, and differential privacy have been extensively studied, functional encryption-based solutions are still less investigated. In this paper, we provide a summary and systematization of privacy-preserving machine learning approaches based on functional encryption. Our analysis assessed the extent to which previous work has addressed the PPML using FE and identified key weaknesses in this area. Additionally, our analysis demonstrates that FE-based approaches could significantly contribute to the PPML issues, but there exist challenges that should be addressed to achieve practical solutions. We hope that this effort paves the path for the research community to investigate this emerging yet important topic.
\bibliography{paper}
\bibliographystyle{acm}
\end{document}